\documentstyle[prd,aps,floats]{revtex}

\newcommand{\be}{\begin{equation}}
\newcommand{\ee}{\end{equation}}
\newcommand{\bea}{\begin{eqnarray}}
\newcommand{\beas}{\begin{eqnarray*}}
\newcommand{\eea}{\end{eqnarray}}
\newcommand{\eeas}{\end{eqnarray*}}
\newcommand{\ba}{\begin{array}}
\newcommand{\ea}{\end{array}}
\newcommand{\bi}{\begin{itemize}}
\newcommand{\ei}{\end{itemize}}
\newcommand{\ben}{\begin{enumerate}}
\newcommand{\een}{\end{enumerate}}

\begin{document}
\draft
\twocolumn[\hsize\textwidth\columnwidth\hsize\csname 
@twocolumnfalse\endcsname

\title{A dynamical stabilization of the radion potential} 
\author{Anupam  Mazumdar$^a$ and A. P\'erez-Lorenzana$^{a,b}$}
\address{
$^a$ The Abdus Salam International Centre for Theoretical Physics, I-34100,
Trieste, Italy \\
$^b$ Departamento de F\'{\i}sica,
Centro de Investigaci\'on y de Estudios Avanzados del I.P.N.\\
Apdo. Post. 14-740, 07000, M\'exico, D.F., M\'exico}
\maketitle

\begin{abstract}
Large extra dimensional theories attempt to solve the hierarchy
problem by assuming that the fundamental scale of the theory 
is at the electroweak scale.
This requires the size of the extra dimensions to be stabilized at 
a scale which is determined by the effective 
four dimensional Planck mass and the 
number of extra dimensions. In this paper we concentrate 
upon the dynamical reasons to stabilize them by providing a running mass 
to the radion field.  We show that it is possible to maintain the  
size of the extra dimensions once it is stabilized throughout the 
dynamics of inflation.
\end{abstract}

\vspace*{-0.6truecm}
\vskip2pc]


\section{introduction} 

Longstanding hierarchy problem, meaning  why the Higgs field has
a mass $m_{\rm H} \sim 1$ TeV, and not the Planck 
mass $M_{\rm p} \sim 10^{18}$GeV as suggested by 
renormalization corrections,
finds a natural solution in supersymmetry where the stability of 
the Higgs mass is provided by the symmetry between bosons and 
fermions. There, the loop contributions from both the sectors 
cancel the quadratic divergences to all orders  keeping the Higgs
mass at the  electroweak scale. So far this has been the preferred solution. 
Recently, however, 
this problem has been addressed without invoking this supersymmetry, 
but, by recognizing the  fundamental scale to be
the higher dimensional Planck mass, $M_*$. In this case $M_{\ast}$ takes 
a value close to the electroweak scale~\cite{nima0,early}.
The four dimensional Planck mass is then obtained via dimensional 
reduction by assuming the extra dimensions are compact. 
Thus, the volume of the extra dimensions $V_d$; the effective four
dimensional Planck mass, and, the fundamental scale are all 
related to each other by a simple relationship
 \be 
 M_{\rm p}^2 = M_*^{2+d} V_d \,,
 \ee
where $d$ is the number of extra dimensions. This automatically determines the
present common size of all the extra dimensions at $b_0$. For two extra
dimensions,  and, $M_*=1$ TeV, the required size is of the order of 
$0.2$mm \cite{exp1,exp2}. This suggests that the extra dimensions are 
indeed  quite large and thus require a trapping potential which can 
stabilize them at their present size. Usually this problem is known 
as a radion stabilization problem, and in fact its stabilization is 
reminiscence to the  problem of moduli and dilaton stabilization in 
low energy effective string and M-theories~\cite{dine}.  
Notice, that in the models with large extra dimensions the
Standard Model particles are trapped in a four dimensional hypersurface 
(a$3$-brane), thus,  they are not allowed to propagate in the bulk.  
However, it is generically assumed that besides gravity, the Standard Model 
singlets may propagate in the bulk.  Among them the  inflaton can be 
a candidate, which is less favored to be a brane  
field~\cite{lyth,linde,abdel}.

It is also  quite unnatural to think that the size of the extra 
dimensions can be fixed during the  evolution of the Universe
right at the electroweak scale. In fact, just
at the electroweak scale the size of the Universe is given 
by~$\sim 10^{-16}$cm. 
Thus, in order to solve the hierarchy we need to inflate the size of the 
extra dimensions \cite{lyth,linde,abdel,anupam,kaloper,dvali,cline}. This 
also tells us that the effective four dimensional theory which 
can be obtained by dimensional reduction must have a new scalar entity, 
which is known as a radion.  The radion also corresponds to the size of the 
extra dimensions as mentioned earlier. For simplicity, we just take this 
radion to be homogeneous and treat it only as a zeroth mode. 
It has been noticed
that the radion has a very light mass $m_{\rm r} \approx 10^{-2}$eV
for two large extra dimensions~\cite{march}.

Now the prime question that appears is how do we stabilize the 
large extra dimensions while keeping all the virtues and  
predictions of the big bang and the inflationary cosmology. This
is a non-trivial issue, however, we can get some insight from the fact
that the expansion in the extra dimensions inevitably introduces
expansion in the four dimensions~\cite{kaloper}. 
This still does not guarantee that the extra dimensions can  
stabilize themselves after expanding their size to the millimeter scale.  
The other important factor which one has to keep in
mind is that this expansion must also provide adequate scale invariant
density perturbation to form the structure formation of the Universe.
In fact the theory of large extra dimensions naturally sets the cut-off
for any mass scale, hence, at most the inflating fields in four dimensions
can acquire a mass $\sim {\rm TeV}$. So keeping all these constraints 
in mind, it is not so easy task to construct a simple model for the early 
Universe. There have been many attempts in this direction, in 
Ref.~\cite{kaloper}, it has been proposed that inflation occurs before
the stabilization of the extra dimensions with the radion field  
playing the role of inflaton. In Ref.~\cite{dvali}, it has been suggested
that the initial displacement of the brane from the stable point can 
lead to inflation in $3$ spatial dimensions.

In this paper we raise the issue of stabilizing the extra dimensions at
a millimeter scale with extremely light radion mass. We argue that this  is
possible provided there exists a radion potential which has a
minimum corresponding to the size of the large extra dimensions. We then show
that it is possible to maintain their large size throughout the dynamics of
inflation. This is the main goal of the present work. In the next section we
set-up our frame work and discuss the initial phase of radion dominated
inflation and the stabilization  mechanism. In the subsequent sections we
describe the radion dynamics, density perturbations, and, reheating of the
Universe. In the last section we briefly summarize our scenario.

\section{Extra dimensions from the point of view of four dimensions}

Let us begin by considering that there exists a scalar field in the bulk.
This scalar field essentially plays a double role; providing 
effective mass to the radion during inflation, and, also acting as 
an inflaton which essentially reheats the Universe before 
nucleosynthesis. Since, the actual mass of the radion is very small,
for instance, for two large extra dimensions it is 
$\sim {\cal O}(10^{-2})$ eV, we can not expect the radion field 
to reheat the Universe to the extent that we obtain 
from a bulk scalar field/fields. On the other hand, we require the bulk
scalar fields in order to produce density
perturbations during inflation~\cite{abdel}. 
Following this wisdom, we invoke 
scalar fields $\chi_{i}$, living in the bulk along with gravity.  
From the point of view of four dimensions the extra dimensions are 
assumed to be compactified on a $d$ dimensional Ricci flat manifold with a 
radii $b(t)$, which have a minimum at $b_0$. The 
higher dimensional metric then reads
 \begin{eqnarray}
 ds^2 = dt^2 -a^2(t) dx^2 -b^2(t)dy^2\,,
 \end{eqnarray}
where $x$ denotes the three spatial dimensions, and $y$ 
collectively denote the extra dimensions. The scale factor of the four 
dimensional space-time is denoted by $a(t)$. After dimensional 
reduction the effective four dimensional action reads \cite{kolb,berkin}
 \begin{eqnarray}
 \label{action}
 S=\int d^4x \sqrt{-g}\left[ -\frac{M_{\rm p}^2}{16\pi}R + \frac{1}{2}
 \partial_{\mu} \sigma \partial^{\mu}\sigma -U(\sigma)\right.\, \nonumber \\ 
 \left.+\frac{1}{2}g^{\mu \nu}\partial_\mu \chi_{i}\partial^\mu \chi_{i}
 -\exp(-d\sigma/\sigma_0)V(\chi_{i}) \right]\,.
 \end{eqnarray}
The four dimensional metric is then determined by $g_{\mu \nu}$, where 
$\mu, \nu =0,1,2,3$. At this moment we are not so specific about 
any matter potential $V(\chi_{i})$, we shall keep the
discussion general until the next section, where we specify a hybrid
inflationary model in order to produce the required density perturbations. 
The radion field is expressed as $\sigma (t)$, which
can be written in terms of the radii of the extra dimensions
\begin{eqnarray}
\label{radion0}
\sigma(t) = \sigma_0 \ln \left[ \frac{b(t)}{b_0}\right]\,, \quad
\sigma_0 =
\left[\frac{d(d+2)M_{\rm p}^2}{16 \pi}\right]^{1/2} \,.
\end{eqnarray}

From the above equation, it is evident that $\sigma_0$ is proportional 
to the four dimensional Planck mass. For $b(t) \sim ({\rm TeV})^{-1}$,
and, $b_0 \sim 1$mm, the modulous of the 
radion field takes a very large initial value. The radion field   
has a potential which is given by $U(\sigma)$ in Eq.~(\ref{action}). 
In this paper we shall not shed any light upon its origin, rather 
we shall assume that such a potential with a minimum is essential 
to stabilize the large extra dimensions. It has been well known that there
can be at least three sources which might contribute to $U(\sigma)$
such as a cosmological constant in $4+d$ dimensional theory. Second 
source might be the curvature of the extra dimensions. However, 
such a contribution can be made zero if the extra dimensions are compactified
on a torus, and, the third possibility could be due to the Casimir force.
All these known possibilities can only help to stabilize the 
extra dimensions close to the electroweak scale and thus they do not serve our
purpose in solving the hierarchy problem. New avenues could be opened
if the extra dimensions were not commutative and a Casimir repulsive force   
could be generated due to a non-trivial topology of the extra 
dimensions~\cite{shahin}. Even then we require to fine tune the 
non-commutative parameter and thus again it is far from convincing. At large 
there is no proper understanding of how such a non-trivial potential 
for the radion field can be generated, and here we do not make any suggestion 
in this direction. Inspite of having a minimum in the radion potential,
it is still a model dependent issue to stabilize the radion around its 
minimum. However, as we feel that the 
issue of radion stabilization is interwined with that of inflation,
and, generation of density perturbation in four dimensions,  it is 
still required to have a simple model where one can address
all these issues together in a single framework.

In this paper we provide a dynamical argument which suggests 
that the radion field can have a running mass during the 
inflationary epoch. When the radion field approaches a 
critical value, it decouples from the rest of the inflationary
dynamics and rolls down the potential to reach the minimum. 
When inflation finally comes to an end, the running
mass of the radion field settles down to its bare mass which 
can be as small as eVs.

\section{Radion dynamics and Inflation}

As we have discussed very briefly in our last section, the radion field 
plays a very significant role in the early Universe, in this paper we
devise a simple scheme which provides a running mass to the radion field. 
This comes quite naturally if there exists a scalar field in the bulk, which
upon dimensional reduction couples to the radion field as shown in the 
effective four dimensional action Eq.~(\ref{action}). 
If we assume that the scalar field $\sigma$, and, at least one of the 
bulk fields; say $\chi$ is rolling down the potential, then
the slow-roll equations yield
 \begin{eqnarray}
 \label{slowr1}
 3H\dot \sigma &\approx &\frac{d}{\sigma_0} e^{-d\sigma/\sigma_0} 
 V(\chi) \,, \\
 \label{slowr2}
 3H\dot \chi &\approx & -e^{-d\sigma/\sigma_0} V^{\prime}(\chi) \,, \\
 \label{slowr3}
 H^2 &\approx & \frac{8\pi}{3M_{\rm p}^2}\left[e^{-d\sigma/\sigma_0}
 V(\chi)+ U(\sigma)\right]\,,
 \end{eqnarray}
where the dot denotes time derivative and prime denotes the derivative with
respect to $\chi$ field. Notice, here we have assumed 
that among many bulk fields only a single field which we call
inflaton is taking part in a slow-roll inflation while the rest of them  are 
trapped in some false vacuum. 
Thus, the potential $V(\chi)$, evaluated at those points, has become
an effective function of $\chi$ alone, however, the bulk of the potential
is coming from the false vacuum.  
On the other hand, the form of $U(\sigma)$ must be such that
there exists a global minimum at $\sigma=0$. For simplicity we can model 
the potential to be $\sim m^2_{\rm r}\sigma^2$. Eventhough the absolute
value of the radion field is extremely large, the radion potential 
$U(\sigma)$ is subdominant compared to the first term 
in the total potential in the above Eq.~(\ref{slowr3})
due to a very small radion mass, which is of the order of eV.

Thus, the effective potential is dominated by the exponential term at 
the beginning of inflation. Notice that eventhough $|\sigma| \approx 34
\sigma_0$ during the onset of inflation, the contribution coming from 
$V(\chi) \sim M_*^2M_{\rm p}^2$, hence, the total energy density
is still bounded by the upper limit $\sim  M_{\rm p}^4$ in our world.
From Eq.~(\ref{radion0}), it is quite evident that in order to obtain 
$b_0$ as large as $1$ mm, the extra dimensions need to grow at least
$35$ e-foldings in size. As we have discussed, if we assume that inflation 
proceeds in $\sigma$ direction, then, from Eqs.~(\ref{slowr1}) and
(\ref{slowr3}), we obtain a simple relation between the expansion rates
of the observable and the extra dimensions~\cite{anupam,berkin}. Which yields 
 \begin{eqnarray}
 \label{expand}
 \frac{d+2}{2}\frac{\dot b}{b} = \frac{\dot a}{a}\,.
 \end{eqnarray}
The set of slow-roll equations satisfy power law inflation in the observable
world for arbitrary number of extra dimensions. The scale factor is given by
 \be
 \label{sol1}
 a(t) \sim {t}^{(d+2)/d}\,,
 \ee
however, during this period the extra dimensions grow much slower as
 \be
 \label{sol2}
 \label{radion1}
 b(t) \sim {t}^{2/d}\,.
 \ee
It is noticeable that the extra dimensions do not inflate for $d\geq 2$
extra dimensions. From the point of view of four dimensions the radion
field inflates our Universe and during this process it rolls down the 
exponential potential to reach the critical value $|\sigma_0|$. About
this point the exponential term in front of $V(\chi)$ in 
Eq.~(\ref{slowr3}), becomes weaker, and this helps the effective
mass of the radion field to catch up with the Hubble expansion
 \be
 \label{mass}
 m^2_{\rm r, eff} \approx m_{\rm r}^2+ \frac{V(\chi)}{\sigma_0^2} \sim 
 {\cal O}(1)H^2\,.
 \ee
Here again, we have neglected the contribution from $U(\sigma)$ compared
to $V(\chi)$, because the bare mass of the radion field
which is of the order of electroweak scale is much smaller than
the Hubble parameter. This is also the reason why we have approximated 
the effective mass to that of the Hubble parameter in the above equation 
Eq.~(\ref{mass}). At this point the radion field can not anymore support
inflation. However, inflation is still continuing, but, now due to the 
vacuum expectation value of the matter fields. This is not an assumption,
but, as we shall see later, it is very important to maintain the mass of the 
radion of the order of Hubble expansion while not disturbing the effective
mass parameters of $\chi$ field. The simplest way to realize inflation 
is from the vacuum dominated phase rather than the slow-roll 
motion of all $\chi_i$ fields. 
This can be satisfied if we assume hybrid potential
for $\chi_i$ fields. In fact, in the next section we shall notice that the
necessity to assume hybrid model has a completely different motivation. The
large effective mass for the radion field helps to decouple its dynamics
from the rest of the fields. If $H$ changes slowly, then the 
field $\sigma$, for $m_{\rm eff}\sim H$, approaches the minimum 
configuration, $\sigma=0$,  exponentially fast. 
We remind the readers that the minimum value of $\sigma$
field is provided by the radion potential $U(\sigma)$. The radion field
evolves like
 \be 
 \label{rest}
 \sigma(t) \approx \sigma_0 e^{\left( - {m_{\rm r, eff}^2(H) t/ 3H} \right)} 
 \sim \sigma_0 e^{\left(- {H t / 3}\right)} \, . 
 \ee

In order to stabilize the radion field we have to ensure that the radion
field settles down to the bottom of the potential much faster before the
end of vacuum dominated inflation. In the standard hybrid model this
phase ends with a phase transition when the
effective mass of the subsidiary field which triggers the phase 
transition becomes imaginary. This takes place when
$\tau \approx 3 H (\delta \ln \phi)/m^2 \approx 3\cdot 10^{-22}$ sec, where 
$m$ is the mass term of the inflaton which we take around 10 GeV. 
Thus, by the time this phase transition takes place, the factor 
$H \tau \sim 10^{8}$ becomes quite large. This leads to an extremely
large suppression to the exponential factor in Eq.~(\ref{rest}). 
In fact the radion field rolls down towards the minimum of $U(\sigma)$
much faster compared to the inflationary time scale, the time taken 
is give by $t\sim1/H\sim 1/M_* = 10^{-30}$ sec.
This is perhaps
the easiest way to freeze the dynamics of the large extra dimensions.

It is noticeable that the effective mass for the radion remains of the 
order of Hubble expansion until the end of vacuum dominated phase. Once
the matter fields begin oscillations and eventually settle down 
at their respective minima via reheating, the effective
mass term for the radion approaches its bare mass $\sim m_{\rm r}$.
However, one might suspect that quantum corrections to the radion mass due to 
its coupling to the matter field, such as Higgs, could be large. Usually 
these corrections must be of the order of the bare mass of the radion 
field due to the Planck mass suppression. There could also be a possible 
contribution that might come from the Higgs vacuum, $v$, which has the 
form $V(v)/M_{\rm P}^2\leq ({\rm eV})^2$. We mention that this contribution 
is directly associated with the cosmological constant problem in the sense that 
it is expected that $V(v)< ({\rm eV})^4$, in order to be consistent 
with current limits~\cite{boomerang}.

\section{Density perturbations}

Any successful inflationary model has to pass through the test of 
density perturbation. As we have noted earlier, the last stage of inflation 
has to be supported by the the vacuum energy density of $V(\chi_i)$, 
whose upper bound is $\sim M_*^2M_{\rm p}^2$. Since, for our purpose, 
the masses of the fields have a natural cut-off scale $\sim M_*$, 
it is extremely difficult to produce adequate density perturbation 
for a simple chaotic inflationary scenario~\cite{lyth,linde}. For this 
reason hybrid model has been chosen because it may overcome a low level 
of density perturbation~\cite{abdel}. In this section we briefly
review the model. Let us consider the $4+d$ dimensional potential 
 \begin{eqnarray}
  V(\hat\chi_1,\hat\chi_2) = {M_*^d\over 4\lambda}
  \left(M_*^2 - {\lambda\over M_*^d} \hat\chi_1^2\right)^2 + 
  {m^2_{\chi_2}\over 2} \hat\chi_2^2 \, \nonumber \\
  + {g^2\over M_*^d} \hat\chi_1^2 \hat\chi_2^2 \,,
 \label{hybp}
 \end{eqnarray}
where $\hat\chi_2$ is the inflaton field, and, $\hat\chi_1$ is the 
subsidiary field which is responsible for the phase transition. Notice,
that the higher dimensional field has a mass dimension $1+d/2$, which leads to
non renormalisable interaction terms. However, the suppression is 
given by the fundamental scale, instead of the four dimensional Planck mass. 
Upon dimensional reduction the effective four dimensional fields, $\chi_i$ 
are related to their higher dimensional relatives by a simple scaling 
 \be 
 \chi_i =  \sqrt{V_d} \hat\chi_i \,.
 \ee
Therefore, eventhough 
the natural initial value for $\hat\chi_i\sim M_*^{1+d/2}$, 
in the  effective four dimensional theory we may have $\chi_i\sim M_{\rm p}$.
It is easy to check that the slow roll conditions for the inflation
breaks down in hybrid model when $\hat\chi_{2c}^2\geq M_*^{2+d}/2g^2$. 
This provides a water fall solution to the fields when 
$\chi_{2c} \approx M_{\rm p}/\sqrt{2}g$. 

Now, in order to perform the density perturbation calculation
we need to know the follow on history of the Universe. Especially, 
we must notice when the observable world should be within the horizon
at the beginning of inflation. This depends on the total number
of e-foldings we have in general. In order to estimate how many e-foldings 
do we get in this model, we make a naive estimation while supposing that 
there are only two extra dimensions.
It is clear that the extra dimensions ought to expand $35$ e-foldings
in order to reach the  millimeter size, this tells us that the three 
spatial dimensions must grow upto $70$ e-foldings during the first phase
of inflation. The second phase of vacuum dominated 
inflation proceeds soon after this initial phase and lasts for at 
least another $40-60$ e-foldings. So, total number of e-foldings 
can be as large as $130$. 
We notice that there are two important differences from the standard 
hybrid inflationary cosmology. First, the inflationary scale is not 
given as usual by the grand unification scale; $\sim 10^{16}$ GeV, but 
in our case the scale is much lower, and it is given by a 
geometric mean; $\sqrt{M_* M_{\rm p}} \sim 10^{12}$ GeV. Second,    the reheat
temperature  of the Universe in our case is extremely low.  Indeed, as we shall
see below reheat temperature is certainly more than MeV, but can be as low as
$T_{\rm r} \sim 100$ MeV,  which for the time being we take as granted while
estimating when the interesting  modes are crossing the present horizon. In
fact, one can easily estimate that what matters is the density perturbations 
produced during the last $43$ e-foldings of inflation~\cite{liddle}. 

In our case, for two large extra dimensions the last $43$ e-foldings 
are supported by  adiabatic perturbations due 
to the slow roll of $\chi_2$ field~\cite{abdel}.
 \begin{eqnarray} 
 {\delta\rho\over \rho} \sim 
 \left({g\over 2 \lambda^{3/2}}\right)\ {M_*^3\over m_{\chi_2}^2 
 M_{\rm p}}.
 \label{drhyb}
 \end{eqnarray}
If we set in the last equation the values; $M_*\sim 10^2\ TeV$, and,
$m_{\chi_2}\sim m\sim 10\ GeV$, we obtain the correct COBE normalization
${\delta\rho/ \rho} \sim \left({g/ 2 \lambda^{3/2}}\right)\times 10^{-5}$,
upto the uncertainties in the coupling constants. So far, we have 
only discussed the adiabatic fluctuations of $\chi_2$ field, in fact
there is another source of density perturbation in our case, which 
is due to the isocurvature fluctuations of the radion field.
Since, we have noticed that the radion field settles down at the bottom
of the potential during inflation with an effective mass 
$\sim {\cal O}(\rm H)$. So, the energy density of the radion field 
is effectively given by its quantum fluctuations, which is quadratic
in nature. The perturbations generated in the density of the radion field
will have a nongaussian feature \cite{linde1,anupam1}. So, in principle
we can expect a mixture of adiabatic and isocurvature fluctuations. 
In this paper we do not delve into the details of the mixed perturbation
scenario.

\section{Reheating}

After hybrid inflation ends the field $\chi_1$  goes to one of its
minima; $\chi_{1\pm}=\pm M_*^{3/2}/\sqrt{\lambda}$.
The inflaton field gets a mass contribution; $\sim g^2M_*^2/\lambda$, 
which dominates over $m_{\chi_2}$. Therefore, by using
the same set of parameters we have introduced to explain density 
perturbations, we can estimate that the physical mass of the oscillating 
inflaton can be about an order of magnitude below $M_*$.
Thus, the decay of $\chi_1$ into Higgs fields; $\phi$, is
allowed for typical Higgs field masses in the range $100-200$ GeV. 
Let us stress that this process will take place only within the brane, 
which marks the departure from the former Kaluza Klein  (KK) theories,
where the production of matter via inflaton decays occurs everywhere. 

The reheating  temperature has been estimated in Ref.~\cite{abdel}. Here we
only recollect the main ideas. First, we calculate the decay rate 
of the inflaton field into two Higgses.  That gives
 \be 
 \Gamma_{\chi_1\rightarrow \phi \phi}\sim 
 {M_*^4\over 32 \pi M_{\rm p}^2 m_{\chi_1}} \,.
 \label{g1}
 \ee
With an inflaton mass $m_{\chi_1} \sim 0.1~M_*$, we obtain  
the reheating temperature to be; $T_{\rm r} \geq 100$ MeV. 
As it is expected, this temperature is above the required temperature
for which the big bang nucleosynthesis can successfully take place. 
It is worth mentioning that bulk reheating is much less efficient than the 
brane reheating, because the production of KK gravitons; $g_{KK}$, is very 
inefficient due to the Planck mass suppressed couplings, and,
KK number conservation~\cite{abdel} which forbids kinematically the 
decay of zero mode inflaton into graviton, or, radion, since, the
process has to have an inflaton mode as a final product.
One may also wonder upon the radion decay. In fact radion can also decay into
two light fermions, through the coupling $(\sigma/\sigma_0) m_\psi\bar\psi
\psi$, however, its decay rate is again Planck mass suppressed. 
Nevertheless, we mention that 
the last e-foldings of inflation, and entropy production during reheating 
is sufficient enough to substantially dilute their number density.

\section{Conclusion}

We conclude our paper by summarizing our main results. We have 
noticed that non-trivial radion dynamics can help to stabilize
the size of the large extra dimensions. In fact the radion mass is
running throughout the inflationary evolution. The initial phase
of inflation is governed by the exponential radion potential. The
radion field rolls down the exponential potential and its effective mass
becomes of the order of the Hubble expansion. When this happens,
the dynamics of the radion field decouples from the rest of the matter
fields. However, inflation continues with the help of vacuum dominated
phase of the hybrid inflationary model. During inflation, the radion field 
rolls down within one Hubble time to its minimum, which helps to
stabilize the size of the extra dimensions. Once, inflation comes 
to an end via the phase transition in the hybrid model, the effective
radion mass drops down to its bare value which is $\sim 10^{-2}$ eV
for the two large extra dimensions. In this model we require last 
$43$ e-foldings of inflation in order to match the density perturbations
from the COBE data. We have shown that the adiabatic fluctuations 
of $\chi_1$ field can produce the adequate density perturbations. 
We have also noticed that there exists another source of
density perturbations. This is due to the quantum fluctuations of 
the radion field around the bottom of the potential. This kind
of density perturbations leads to nongaussian spectrum. In fact 
we would expect to have mixture of both adiabatic and isocurvature
fluctuations. It will be interesting to study the details of the 
density fluctuations, which might lead to a definitive prediction of
our model.

\acknowledgements
A. M. acknowledges the support of {\bf The Early Universe network} 
HPRN-CT-2000-00152.


\end{document}